\documentclass[twoside]{article}
\usepackage{graphicx,amssymb,mathrsfs,amsmath,latexsym,amsfonts,amsthm}
\input vatola.sty \input cyracc.def

\renewcommand{\baselinestretch}{1.06} 
\catcode`@=11 \long\def\@makefntext#1{\noindent #1}
\newskip\tabcentering \tabcentering=1000pt plus 1000pt minus 1000pt
\def\REF#1{\par\hangindent\parindent\indent\llap{#1\enspace}\ignorespaces}
\def\MCH#1#2{\setbox0=\hbox{\raise#1\hbox{#2}}\smash{\box0}}

\let\@oddfoot\@empty  \let\@evenfoot\@empty
\TagsOnRight

\def\@evenfoot{\footnotesize{\thepage\hfill}}\def\@oddfoot{\footnotesize{\hfill\thepage}}
\def\@evenhead{\hfill\footnotesize{{\it YAO Yong}}}
\def\@oddhead{{\footnotesize
{\it Successive difference substitution based on column stochastic
matrix} \hfill}}

\def\sec#1{\vspace{2mm}\noindent{{\bf #1}}\vspace{0.5mm}}
\def\subsec#1{\vspace{1mm}\noindent{{\bf #1}}\vspace{0.5mm}} 
\def\th#1{\vspace{1mm}\noindent{\bf #1}\quad} 
\def\pf#1{\vspace{1mm}\noindent{\it #1}\quad}

\def\leq{\leqslant}
\def\geq{\geqslant}
\def\R{{\Bbb R}}  \def\N{{\Bbb N}}  \def\Q{{\Bbb Q}}   
  \def\wj{\end{document}}  \newsymbol\wjzhml 203F \def\no{\noindent}

\begin{document}
\abovedisplayskip=3pt plus 1pt minus 2pt 
\belowdisplayskip=3pt plus 1pt minus 2pt 

\

\textwidth=145truemm \textheight=218truemm \thispagestyle{empty}
\renewcommand{\baselinestretch}{0.9}\baselineskip 9pt
\begin{picture}(0,0)
\put(-16,0){\vbox{\hbox{\footnotesize\it Science in China $:$
Mathematics} \hbox {\footnotesize 2010 Vol. 40 No. 3: 251--264}
\hbox{\footnotesize www.scichina.com}}}
\end{picture}

\vspace{8true mm}

\renewcommand{\baselinestretch}{1.6}\baselineskip 15pt

\noindent{\LARGE\bf Successive Difference Substitution Based on\\
Column Stochastic Matrix and Mechanical\\ Decision for Positive
 Semi-definite Forms}
\vspace{0.5 true cm}

\noindent{\normalsize\sf  YAO Yong 
\footnotetext{\baselineskip 10pt Received August 8, 2008;
accepted October 27, 2009\\
This work was partially supported by the National Key Basic Research
Project of China (Grant No. 2004CB318003) and
the National Natural Science Foundation of China(Grant No. 90418041, 10901116)}}

\vspace{0.2 true cm}
\renewcommand{\baselinestretch}{1.3}\baselineskip 12pt
\noindent{\footnotesize\rm  Chengdu Institute of Computer
Applications,
 Chinese Academy of Sciences, Chengdu 610041, China\\
(email: yaoyong@casit.ac.cn, yaoyong@yahoo.cn)
\vspace{4mm}}%

\baselineskip 12pt \renewcommand{\baselinestretch}{1.0}
\noindent{{\bf Abstract}\small\hspace{2.8mm}

The theory part of this paper is sketched as follows. Based on
column stochastic average matrix $T_n$ selected as a basic
substitution matrix, the method of advanced successive difference
substitution is established. Then, a set of necessary and sufficient
conditions for deciding positive semi-definite form on $\R^n_+$ is
derived from this method. And furthermore, it is proved that the
sequence of SDS sets of a positive definite form is positively
terminating.

Worked out according to these results, the Maple program TSDS3 not
only automatically proves the polynomial inequalities, but also
outputs counter examples for the false. Sometimes TSDS3 does not
halt, but it is very useful by experimenting on so many examples.


\vspace{1mm} \no{\footnotesize{\bf Keywords:\hspace{2mm} Successive
Difference Substitution, Sequence of SDS sets, Termination, Positive
Semi-definite, Machine decision}}

\no{\footnotesize{\bf MSC(2000):\hspace{2mm} 68T15, 26D05 }
\vspace{2mm} \baselineskip 15pt
\renewcommand{\baselinestretch}{1.10}
\parindent=16pt  \parskip=2mm
\rm\normalsize\rm

\sec{1\quad Introduction}

A traditional method which is now called difference substitution
(DS) can be frequently found and applied to the proof of
inequalities with symmetric form. However, this method does not
consider asymmetric forms; it is also invalid for the problems
failed with the first substitution. To remedy that, Yang defined a
general concept, difference substitution([1][2]) , for both
asymmetric and symmetric forms, and also introduced the method
called successive difference substitution (SDS) which is the key to
the latter. According to these, Yang designed a heuristic Maple
program SDS, which has been applied to prove a great many polynomial
inequalities with more variables and higher degrees, some of which
are still considered very hard with other computer algebra methods.

At the 2nd Summer School on Symbolic Computation (Beijing, China,
July 15-22, 2006), Yang put forward an open problem$^{[3]}$ on the
termination of successive difference substitution. That is, which
kind of polynomials can be solved by SDS and which kind can not
(here ``can be solved '' means that SDS can prove the polynomial
inequality $f\geq 0$)? This problem, regarded as extremely
difficult, has been spread widely. Furthermore, how to apply the
method to deciding whether an inequality is false? Like traditional
method, Yang's did not take it into account. In this paper, the
method is improved on both counts. Thus, both in theory and
application scope, the method of difference substitution is much
stronger than before.

The traditional method of difference substitution chooses the upper
triangular matrix $A_n$ as the basic substitution matrix. However,
as $m\rightarrow \infty$, $A_n^m$ does not converge. Here we can
choose the matrix $K\ (q_i>0)$ as the basic substitution matrix. It
is a pity that $K^m$ does not always converge when $m\rightarrow
\infty$. These make the discussion of termination of the sequence of
SDS sets much more difficult. So, finally we select $T_n$ as the
basic substitution matrix in order to overcome the termination
problem more easily.

$A_n$, $K$ and $T_n$ are respectively denoted by,
\begin{eqnarray}
A_{n}= \left[
\begin{array}{cccc}
1 & 1 & \cdots & 1 \\
0 & 1 & \ddots & \vdots \\
\vdots& \ddots &\ddots &1 \\
0 & \cdots & 0 & 1
\end{array}
\right], K =\left[
\begin{array}{cccc}
q_1 & q_2 & \cdots & q_n \\
0  & q_2  & \ddots & \vdots \\
\vdots& \ddots &\ddots &q_n \\
0  & \cdots & 0  & q_n
\end{array}
\right], T_{n}= \left[
\begin{array}{cccc}
 1 & \frac{1}{2}  & \cdots & \frac{1}{n}\\
 0 & \frac{1}{2}  & \ddots & \vdots \\
 \vdots& \ddots &\ddots &\frac{1}{n} \\
 0 & \cdots & 0  & \frac{1}{n} \\
\end{array}
\right].
\end{eqnarray}

It is easily to prove that $\lim \limits_{m\rightarrow \infty
}T_n^m$ exists. And $T_n$ has a better property: Let ${\rm PT}_n$ be
the set consisting of matrices transformed by swapping any two rows
of $T_n$, which comprises $n!$ elements. Choose countable infinite
matrices from ${\rm PT}_n$ at random (repeats allowed), then the
product of these infinite matrices will converge to a column
stochastic matrix with rank 1. More generally, there is:

\th{Main Lemma.$^{[4][5]}$}\ \  {\it Let $\Omega$ be a set
consisting of finite $n\times n$  column stochastic mean matrices
{\em (Def 2.1)}. $B_m\in \Omega$. Then here exists nonnegative real
numbers\ $b_1,\cdots,b_n$, $b_1+b_2+\cdots +b_n=1$, such that}
$$\prod \limits_{m=1}^{\infty} B_m=
\left[
\begin{array}{cccc}
b_1 & b_1 & \cdots & b_1 \\
b_2 & b_2 & \cdots & b_2 \\
\vdots& \vdots &\vdots &\vdots \\
b_n & b_n &\cdots & b_n
\end{array}
\right].$$

Main Lemma is not a new result. It has been found by many
researchers[4][5]. However, we will give our proof in order to
explain the action of Main Lemma to the main results of this paper.

\th{Definition 1.1.}\ {\it The form {\em (namely, homogeneous
polynomials)} $f(x_1,x_2,\cdots,x_n)\in \R[x_1,\allowbreak
x_2,\cdots,x_n]$ is called positive semi-definite on $\R^n_+$ {\em
(PSD)}, if
$$(\forall(x_1,x_2,\cdots,x_n)\in \R^n_+)\
f(x_1,x_2,\cdots,x_n)\geq 0,$$ where $\R^n_+=\{(x_1,x_2,\cdots,x_n)|
x_1\geq 0,x_2\geq 0\,\cdots,x_n\geq 0\}$. If $f>0$ for all
$(x_1,x_2,\cdots,x_n)\neq(0,0,\cdots,0)$, then $f$ is called a
positive definite form on $\R^n_+$} (briefly, a PD).

Here is the main results in this paper:

\th{Theorem 1.}\ {\it $1^\circ$ The form $f\in $ PSD iff the
sequence of SDS sets {\em (Def 3.4)} of $f$ is positively
terminating {\em (Def 3.6)} or not terminating.

$2^\circ$ The form $f\notin$ PSD iff the sequence of SDS sets of $f$
is negatively terminating {\em (Def 3.7)}.

$3^\circ$  If the form $f\in $ PD, then its sequence of SDS sets is
positively terminating.}

The paper is organized as follows. Section 2 gives the basic
properties of column stochastic mean matrix and the proof to the
main Lemma. Section 3 introduces some preliminary notions of
difference substitution. The proof of main results and the
establishment of algorithm is presented in section 4 and the
comparison with P\`{o}lya theorem in section 5.

\sec{2\quad The Column Stochastic Mean Matrix and the Proof for Main
Lemma}

\subsec{2.1\quad The column stochastic mean matrix}

A column stochastic matrix is a square matrix whose columns consist
of nonnegative real numbers whose sum is 1. The column stochastic
matrix has a very important property, that is, let
$e=(1,1,\cdots,1)$, and $A=[a_{i,j}]$ is a column stochastic matrix,
then there is $eA=e$.

\th{Lemma 2.1}\ {\it Let $\alpha =(x_1,\ x_2,\ \cdots,\ x_n)\in
\R_+^n$, where $x_1+\cdots+x_n=1$. And $\beta
=(b_1,b_2,\cdots,b_n)\in \R_+^n$, where the maximal element of
$\beta$ is denoted with $\max(\beta)$, and the minimal element with
$\min(\beta)$ . Then there are

$1^\circ \quad\min(\beta)\leq x_1b_1+x_2b_2+\cdots+x_nb_n \leq
\max(\beta).$

$2^\circ$ \ Let the subscripts that $\min(\beta)$ appears on $\beta$
are $s_1,s_2,\cdots,\ s_k\ (1\leq k\leq n)$ (that is
$b_{s_1}=b_{s_2}=\cdots=b_{s_k}=\min(\beta)$), then $\min(\beta)=
x_1b_1+x_2b_2+\cdots+x_nb_n$ iff $x_{s_1}+x_{s_2}+\cdots
+x_{s_k}=1.$

$3^\circ$ \ Let the subscripts that $\max(\beta)$ appears on $\beta$
are $r_1,r_2,\cdots,\ r_t\ (1\leq t\leq n)$, then $\max(\beta)=
x_1b_1+x_2b_2+\cdots+x_nb_n$ iff $x_{r_1}+x_{r_2}+\cdots
+x_{r_t}=1.$

$4^\circ$ \ $\min(\beta)=\max(\beta)$ iff all the elements of
$\beta$ are equal.}

The proof for Lemma 2.1 is obviously and is omitted. Next, there be
the definition of column stochastic mean matrix.

\th{Definition 2.1}\ {\it Let $Q$ be a $n\times n$ column stochastic
matrix, and $\beta=(b_1,b_2,\cdots,b_n)$ be any set of nonnegative
real numbers that are not totally equal. If
$$[\min(\beta Q),\ \max(\beta Q)] \subset [\min(\beta),\max(\beta)].$$
holds, then $Q$ is a column stochastic mean matrix.}

The condition of Definition 2.1 is very strict. Consider the
following two matrices,
\begin{eqnarray}
Q_1= \left[
\begin{array}{ccc}
\frac{1}{2} & 0 &0 \\
\frac{1}{2}& \frac{1}{2}&0 \\
0 &\frac{1}{2} &1
\end{array}
\right],\ Q_2= \left[
\begin{array}{ccc}
\frac{1}{2} & 0 &\frac{1}{2}\\
\frac{1}{2}& \frac{1}{2}&0 \\
0 &\frac{1}{2} &\frac{1}{2}
\end{array}
\right],
\end{eqnarray}
in which $Q_1$ is not a column stochastic mean matrix. Because for a
vector $\beta=(0,0,1)$, we have $[\min(\beta),\ \max(\beta)]=[0,1]$
and $[\min(\beta Q_1),\max(\beta Q_1)]=[0,1]$, which doesn't satisfy
the condition of Definition 2.1. And we can prove that $Q_2$
satisfies the condition of Definition 2.1. Thus there is a natural
question, that is, what condition makes a column stochastic matrix
become a column stochastic mean matrix?

\th{Lemma 2.2.}\  {\it The $n\times n$ column stochastic matrix
$A=[a_{i,j}]$ is a column stochastic mean matrix iff for two
arbitrary columns of $A$
$${\rm Col}(A,j_1)=(a_{1,j_1},\ a_{2,j_1},\cdots,\ a_{n,j_1})^T,\
{\rm Col}(A,j_2)=(a_{1,j_2},\ a_{2,j_2},\cdots,\ a_{n,j_2})^T$$
there be}
$$(a_{1,j_1}a_{1,j_2},\ a_{2,j_1}a_{2,j_2},\cdots,\ a_{n,j_1}a_{n,j_2})\neq (0,0,\cdots,0).$$

\pf{Proof.}\  $\Rightarrow$: (proof by contradiction) Suppose there
are two columns ${\rm Col}(A,j_1),{\rm Col}(A,j_2)$ of the column
stochastic mean matrix $A=[a_{i,j}]$ such that
$$(a_{1,j_1}a_{1,j_2} ,\ a_{2,j_1}a_{2,j_2},\cdots,\ a_{n,j_1}a_{n,j_2})= (0,0,\cdots,0).$$
we might as well suppose that
\begin{eqnarray*}
a_{1,j_1}\neq 0,\cdots,\ a_{t,j_1}\neq 0,\ a_{t+1,j_1}=0, \cdots,\ a_{n,j_1}=0;\\
a_{1,j_2}=0,\cdots,\ a_{k,j_2}= 0,\ a_{k+1,j_2}\neq 0, \cdots,\
a_{n,j_2}\neq 0.
\end{eqnarray*}
where $1\leq t\leq k< n$.

Let $\beta(t)=(\underbrace{1,\cdots,1}\limits_{t},0,\cdots,0)$, then
$[\min(\beta(t)),\max(\beta(t))]=[0,1]$. By Lemma 2.1, there be
$[\min(\beta(t)A),\max(\beta(t)A)]=[0,1],$ a contradiction.

$\Leftarrow$: Let $A=[a_{i,j}]$ be a $n\times n$ column stochastic
matrix, and pick any two columns of it,
$${\rm Col}(A,j_1)=(a_{1,j_1},\ a_{2,j_1},\cdots,\ a_{n,j_1})^T,\
{\rm Col}(A,j_2)=(a_{1,j_2},\ a_{2,j_2},\cdots,\ a_{n,j_2})^T$$
Then, by known, we have
$$(a_{1,j_1}a_{1,j_2} ,\ a_{2,j_1}a_{2,j_2},\cdots,\ a_{n,j_1}a_{n,j_2})\neq (0,0,\cdots,0).$$
That is, at least there is one positive integer $s\ (1\leq s\leq n)$
such that $a_{s,j_1}\neq 0,a_{s,j_2}\neq 0$.

Let $\beta_1=(b_1,b_2,\cdots,b_n)$ be any set of nonnegative real
numbers that are not totally equal. And suppose that the subscripts
that $\min(\beta_1 )$ appears on $\beta_1$ are $s_1,s_2,\cdots,\
s_k\ (1\leq k\leq n)$, and the subscripts that $\max(\beta_1)$
appears on $\beta_1$ are $r_1,r_2,\cdots,\ r_t\ (1\leq t\leq n)$.

Consider the product of $\beta_1$ and the column $j_1,j_2$ of $A$,
and note

 $\beta_{2j_1}=b_1a_{1,j_1}+b_2a_{2,j_1}+\cdots+b_na_{n,j_1},$

 $\beta_{2j_2}=b_1a_{1,j_2}+b_2a_{2,j_2}+\cdots+b_na_{n,j_2}.$\\
Suppose that
$\beta_{2j_1}=\min(\beta_1),\beta_{2j_2}=\max(\beta_1)$, then by
$2^\circ,\ 3^\circ$ of Lemma 2.1 we have

$a_{s_1,j_1}+a_{s_2,j_1}+\cdots+a_{s_k,j_1}=1,$

$a_{r_1,j_2}+a_{r_2,j_2}+\cdots+a_{r_t,j_2}=1.$

On the one hand, the matrix $A$ is column stochastic, so the nonzero
numbers of the column $j_1$ of $A$ only appear in $(a_{s_1,j_1}, \
a_{s_2,j_1},\allowbreak \cdots,\ a_{s_k,j_1})$. Similarly, the
nonzero numbers of the column $j_2$ only appear in $(a_{r_1,j_2},\
a_{r_2,j_2},\ \cdots,\ a_{r_t,j_2})$. But there is at least one
positive integer $s\ (1\leq s\leq n)$ such that $a_{s,j_1}\neq
0,a_{s,j_2}\neq 0$. Thus there is at least one common element both
in $(s_1,s_2,\cdots,\ s_k)$ and $(r_1,r_2,\cdots,\ r_t)$.

On the other hand, because $\min(\beta_1)\neq \max(\beta_1)$, then
elements in $(s_1,s_2,\cdots,\ s_k,\ r_1,\allowbreak r_2,\cdots,\
r_t)$ are distinct. So, there is a contradiction.

The reason for the above contradiction is that we mistakenly suppose
that both $\min(\beta_1)$ and $\max(\beta_1)$ are in $\beta_1A$. So
$\min(\beta_1)$ and $\max(\beta_1)$ can't be in $\beta_1A$ at one
time. And by $1^\circ$ of Lemma 2.1, thus sufficiency holds.

For the contracting intervals, whether the length of a interval will
converge to 0 is very important. So, for a fixed matrix $A$, we
should consider the maximal value of function
\begin{eqnarray}
M_A(X)=\frac{\max(X A)-\min(X A)}{\max(X)-\min(X)},
\end{eqnarray}
 where $X=(x_1,\cdots,x_n),\ x_i\geq 0$ and $\max(X)-\min(X)\neq
0$.

\th{Lemma 2.3.}\ {\it Given a fixed column stochastic mean matrix
$A$. The maximal value of function $M_A(X)$ exists under the
condition $[x_i\geq 0$ ,$\max(X)-\min(X)\neq 0]$, and it is less
than 1.}

\pf{Proof.}\  Firstly, suppose that $\min(X)=0$(otherwise use
$\widetilde{X}= (x_1-\min(X),x_2-\min(X),\cdots,x_n-\min(X))$
instead of $X$). It is easily to prove that
$M_A(\widetilde{X})=M_A(X)$.

Secondly, suppose that $\max(X)=1$(otherwise use
$\widetilde{X}=(x_1/\max(X),x_2/\max(X),\cdots,\\x_n/\max(X))$
instead of $X$). Similarly, we can prove that
$M_A(\widetilde{X})=M_A(X)$.

Thus we only consider the simplified function
\begin{eqnarray*}
M_A(X)=\max(X A)-\min(X A),
\end{eqnarray*}
where $0 \leq x_i\leq 1$ and $\max(X)=1,\min(X)=0$.

Repeatedly use the formulas
$$
\max(x,y)=((x+y)+|x-y|)/2,\  \min(x,y)=((x+y)-|x-y|)/2.
$$
It is easily to find that $M_A(X)$ is continuous on the bounded
closed region $[0 \leq x_i\leq 1,\ \max(X)=1,\min(X)=0]$ , so the
maximal value of it exists and less than 1.

Let $M_A=\max(M_A(X))$, and we call $M_A$ the contraction
coefficient of the column stochastic mean matrix $A$. For a concrete
matrix $A$,\ $M_A$ is easy to solve. Here is an example.

\th{Example 1.} Given
\begin{eqnarray*}
Q= \left[
\begin{array}{ccc}
\frac{1}{3} & 0 &0\\
\frac{1}{3} & \frac{1}{2}&0 \\
\frac{1}{3} &\frac{1}{2} &1
\end{array}
\right],
\end{eqnarray*}
find the contraction coefficient $M_{Q}$ of $Q$.

\pf{Solution.} Suppose that $X=(x_1,x_2,x_3)$, there are six
different possible forms after the simplification of $X$, that is,
$$(0,x_2,1),(1,x_2,0),(x_1,0,1),(x_1,1,0),(0,1,x_3),(1,0,x_3).$$
\begin{eqnarray*}
&&M_{Q}(0,x_2,1)=\max((0,x_2,1)Q)-\min((0,x_2,1)Q)\\
&&\qquad \qquad \qquad=\max((x_2+1)/3,(x_2+1)/2,1)-\min((x_2+1)/3,(x_2+1)/2,1)\\
&&\qquad \qquad \qquad=1-(x_2+1)/3 \leq 2/3\ (\hbox{holds when }
x_2=0).
\end{eqnarray*}

Analogously, it still holds for the other forms. So $M_{Q}=2/3$.

\th{Lemma 2.4.}\ {\it Let $X=(x_1,\ x_2,\ \cdots,\ x_n)$ is a set of
nonnegative real numbers that are not totally equal, and $A$ is a
$n\times n$ column stochastic mean matrix. Note that the length of
interval $[a,b]$ is $d([a,b])$, then}
$$d([\min(XA),\max(XA)])\leq M_A d([\min(X),\max(X)]).$$

The purpose of the contraction coefficient $M_A$ is to denote the
above relation.

\subsec{2.2\ Proof for the Main Lemma}

Main Lemma is not only very important for the proof of the main
results but also very valuable itself.

\th{Main Lemma.$^{[4][5]}$}\ \  {\it Let $\Omega$ be a set
consisting of finite $n\times n$ column stochastic mean matrices.
$B_m\in \Omega$. Then here exists nonnegative real numbers
$b_1,\cdots,b_n$, $b_1+b_2+\cdots +b_n=1$, such that}
$$\prod \limits_{m=1}^{\infty} B_m=
\left[
\begin{array}{cccc}
b_1 & b_1 & \cdots & b_1 \\
b_2 & b_2 & \cdots & b_2 \\
\vdots& \vdots &\vdots &\vdots \\
b_n & b_n &\cdots & b_n
\end{array}
\right].$$

\pf{Proof.} For $\Omega$ is a finite set, then there is a matrix
whose contraction coefficient is maximal in it. Denote the maximal
contraction coefficient with $M_{\Omega}$.

Let the column stochastic mean matrices picked in turn from $\Omega$
are
$$B_1,\ B_2,\ \cdots,\ B_m,\cdots$$

And let $B_1=[a_{i,j}]$, take the first row of $B_1$ as the sequence
of vectors $\beta_1,\beta_2,\cdots,$
\begin{eqnarray*}
&&\beta_1=(a_{1,1},\ a_{1,2},\cdots,\ a_{1,n}),\\
&&\beta_2=(a_{1,1},\ a_{1,2},\cdots,\ a_{1,n})B_2,\\
&&\cdots,\cdots,\\
&&\beta_m=(a_{1,1},\ a_{1,2},\cdots,\ a_{1,n})B_2B_3\cdots B_m,\\
&&\cdots,\cdots.
\end{eqnarray*}

If there is a positive integer $k$ such that
$\min(\beta_k)=\max(\beta_k)$, then by column stochastic, it is
clearly that the sequence $\{\beta_m\}_{m=1}^\infty$ converges to
$(b_1,b_1,\cdots,b_1),\ b_1=\min(\beta_k)$. Otherwise, by Definition
2.1, there is a nested closed set such that
\begin{eqnarray*}
[\min(\beta_1),\max(\beta_1)]\supset [\min(\beta_2),\max(\beta_2)]
\supset \cdots \supset [\min(\beta_m),\max(\beta_m)]\supset \cdots.
\end{eqnarray*}

And by Lemma 2.4, the length of the $m^{th}$ interval is less than
$d([\min(\beta_1),\max(\beta_1)])M_{\Omega}^{m-1}$. Thus, when
$m\rightarrow \infty$, the length of the interval converges to 0.

 According to Squeeze Theorem, the sequence of vectors
 $\beta_1,\beta_2,\cdots$ converges to $(b_1,b_1,\cdots,b_1)$, where
$$b_1=\lim \limits_{m\rightarrow \infty}\min(\beta_m)=
\lim \limits_{m\rightarrow \infty}\max(\beta_m).$$

Analogously, pick the prime vector $\beta_1$ as the $i^{th}$ row of
the matrix $B_1$, then we have the corresponding limit vector which
is denoted as $(b_i,b_i,\cdots,b_i)$. Thus
$$\prod \limits_{m=1}^{\infty} B_m=
\left[
\begin{array}{cccc}
b_1 & b_1 & \cdots & b_1 \\
b_2 & b_2 & \cdots & b_2 \\
\vdots& \vdots &\vdots &\vdots \\
b_n & b_n &\cdots & b_n
\end{array}
\right].$$

\sec{3\quad Preliminary Notions of Difference Substitution}

$n\times n$ upper triangular matrix $T_n$ and $T^{-1}_n$ are
respectively
\begin{eqnarray*}
T_{n}= \left[
\begin{array}{cccc}
 1 & \frac{1}{2}  & \cdots & \frac{1}{n}\\
 0 & \frac{1}{2}  & \ddots & \vdots \\
 \vdots& \ddots &\ddots &\frac{1}{n} \\
 0 & \cdots & 0  & \frac{1}{n} \\
\end{array}
\right],\ T^{-1}_{n}= \left[
\begin{array}{cccc}
1 & -1 & \cdots & 0 \\
0 & 2 & \ddots & \vdots \\
\vdots& \ddots &\ddots &-(n-1) \\
0 & \cdots & 0 & n
\end{array}
\right].
\end{eqnarray*}

Let $S_n$ be a symmetric group of $n$ letters, $\sigma\in S_n$. Then
$P_{\sigma}$ will be the $n \times n$ permutation matrix
corresponding to $\sigma$.

\th{Definition 3.1.}\  {\it $n\times n$ square matrix $B_{\sigma}$
is defined as follows:
$$
B_{\sigma}=P_{\sigma}T_n.
$$
And the set that consists of all $B_{\sigma}$ is denoted by ${\rm
PT}_n$ }(there are $n!$ elements in ${\rm PT}_n$).

The standard simplex on $\R^n_+$ is defined as follows,
$$\Delta_n=\{(x_1,x_2,\cdots,x_n)^T|\ x_1\geq 0,x_2\geq
0,\cdots,x_n\geq 0,\ x_1+x_2+\cdots +x_n=1\}.$$ Let
$X=(x_1,x_2,\cdots,x_n)^T$. Clearly, if $X\in \Delta_n$, then
$B_{\sigma}X\in \Delta_n.$

\th{Definition 3.2.}\ {\it Let $X=(x_1,x_2,\cdots,x_n)^T, \sigma\in
S_n$ Then we set a linear mapping}
$$
\begin{array}{rl}
\psi_{\sigma}:&\Delta_n \rightarrow  \Delta_n,\\
\quad &X \mapsto B_{\sigma}X.
\end{array}
$$

Furthermore, composition of mapping $\psi_{\sigma}$ is respectively
$$\psi_{\sigma_1}\psi_{\sigma_2}\cdots \psi_{\sigma_m}(X)
=\psi_{\sigma_1}(\psi_{\sigma_2}\cdots (\psi_{\sigma_m}(X))\cdots
)=B_{\sigma_1}B_{\sigma_2}\cdots B_{\sigma_m}X,$$ 这里$\sigma_i\in
S_n$.

The image set of $\psi_{\sigma_1}\psi_{\sigma_2}\cdots
\psi_{\sigma_m}$ is denoted by ${\rm
Im}(\psi_{\sigma_1}\psi_{\sigma_2}\cdots \psi_{\sigma_m})$. Let
$P_1,\cdots,P_n$ be points of $\Delta_n$ that are linearly
independent. The simplex with vertices $P_1,\cdots,P_n$ is denoted
by $[P_1,\cdots,P_n]$. Then $\Delta_n=[e_1,e_2,\cdots,e_n]$, where
$e_1=(1,0,\cdots,0)^T, \cdots, e_n=(0,0,\cdots,1)^T$.

Obviously, ${\rm Im}(\psi_{\sigma_1}\psi_{\sigma_2}\cdots
\psi_{\sigma_m})$ is the simplex $\Delta_n
B_{\sigma_1}B_{\sigma_2}\cdots B_{\sigma_m}$. Thus, we briefly note
$${\rm Im}(\psi_{\sigma_1}\psi_{\sigma_2}\cdots
\psi_{\sigma_m})=\Delta_n B_{\sigma_1}B_{\sigma_2}\cdots
B_{\sigma_m}$$

We use $d(S)$ for representing the diameter of simplex $S$, then we
have the following Lemma.

\th{Lemma 3.1.}\ {\it Let $P_1,\cdots,P_n$ be points of $\Delta_n$
that are linearly independent. The simplex with vertices
$P_1,\cdots,P_n$ is denoted by $[P_1,\cdots,P_n]$. Then there are}

$1^\circ$\ $$[P_1,P_2,\cdots,P_n]=\bigcup \limits_{\sigma \in
S_n}[P_1,P_2,\cdots,P_n]B_{\sigma}.$$

$2^\circ$\ $$\frac{d([P_1,P_2,\cdots,P_n])}
{d([P_1,P_2,\cdots,P_n]B_{\sigma})}\leq \frac{n-1}{n}.$$

\pf{Proof.} (Sketch) $1^\circ$  For identity permutation I, we
Consider the simplex
$[P_1,P_2,\cdots,P_n]B_{I}=[P_1,P_2,\cdots,P_n]T_n$, namely,
$$[P_1,\frac{P_1+P_2}{2},\cdots,\frac{P_1+P_2+\cdots+P_n}{n}].$$
Note that $\frac{P_1+P_2+\cdots+P_i}{i}$ is the barycenter of
subsimplex $[P_1,P_2,\cdots,P_i]$, then
$[P_1,P_2,\cdots,P_n]B_{\sigma}$ is exactly the barycentic
subdivision of $[P_1,\cdots,P_n]$.

The result of $2^\circ$ is a elementary property of barycentic
subdivision ([6][7]).

By Lemma 3.1, we immediately have

\th{Lemma 3.2.}\ $1^\circ$
$$\Delta_n=\bigcup \limits_{\sigma_m\in S_n}\cdots \bigcup \limits_{\sigma_2\in S_n}
\bigcup \limits_{\sigma_1\in S_n}(\Delta_n
B_{\sigma_1}B_{\sigma_2}\cdots B_{\sigma_m}).$$

$2^\circ$
$$\frac{d(\Delta_n B_{\sigma_1}B_{\sigma_2}\cdots B_{\sigma_m})}{d(\Delta_n)}\leq (\frac{n-1}{n})^m.$$

\th{Definition 3.3.}\ {\it $f\in \R[x_1,\cdots,x_n],
X=(x_1,\cdots,x_n)^T$,\ $S_n$ be the symmetric group on $n$ letters.
we define the set
$${\rm SDS}^{(m)}(f)=\bigcup \limits_{\sigma_m\in S_n}\cdots \bigcup \limits_{\sigma_2\in S_n}
\bigcup \limits_{\sigma_1\in S_n}f(B_{\sigma_1}B_{\sigma_2}\cdots
B_{\sigma_m}X),$$  which is called the set of difference
substitution with degree $m$ of the form $f$}.

Generally, ${\rm SDS}^{(m)}(f)$ is constructed from $(n!)^m$
polynomials. Definition 3.3 is only in appearance different from the
corresponding one of [1]. So it belongs to Yang's contribution.

Now it's time to define the sequence of SDS sets.

\th{Definition 3.4.}\ {\it Given the form $f\in
\R[x_1,x_2,\cdots,x_n]$, we define the sequence of sets \allowbreak
$\{{\rm SDS}^{(m)}(f)\}_{m=1}^{\infty}$ {\em (the sequence of SDS
sets)} as follows}
\begin{equation}
\{{\rm SDS}^{(m)}(f)\}_{m=1}^{\infty}={\rm SDS}(f),\ {\rm
SDS}^{(2)}(f),\ \cdots.
\end{equation}

We will define the termination of this sequence, which is directly
related to the positive semi-definite property of form $f$

Let $\alpha=(\alpha_1,\alpha_2,\cdots,\alpha_n)\in \N^n$,\ and let
$|\alpha|=\alpha_1+\alpha_2+\cdots+\alpha_n$. Then we write a form
$f$ with degree $d$ as
\begin{equation}
f(x_1,x_2,\cdots,x_n)=\sum\limits_{|\alpha|=d}C_{\alpha}x^{\alpha_1}_1
x^{\alpha_2}_2\cdots x^{\alpha_n}_n,\ C_{\alpha}\in \R.
\end{equation}

\th{Definition 3.5.}\ {\it The form $f$ is called trivially positive
if the coefficients $C_{\alpha}$ of every terms
$x^{\alpha_1}_1x^{\alpha_2}_2\cdots x^{\alpha_n}_n$ in $f$ are
nonnegative. If $f(1,1,\cdots,1)<0$, then $f$ is trivially
negative.}

Obviously, if the form $f$ is trivially positive, then $f\in$PSD. If
the form $f$ is trivially negative, then $f\notin$ PSD.

Then there are some definitions on termination of $\{{\rm
SDS}^{(m)}(f)\}_{m=1}^{\infty}$.

\th{Definition 3.6.}\ {\it Given a form $f$, if there is a positive
integer $k$\ such that every element of the set ${\rm SDS}^{(k)}(f)$
is trivially positive, the sequence of sets $\{{\rm
SDS}^{(m)}(f)\}_{m=1}^{\infty}$ is called positively terminating.}

\th{Definition 3.7.}\ {\it Given a form $f$, if there is a positive
integer $k$ and a form $g$ such that $g\in {\rm SDS}^{(k)}(f)$ and
$g$ is trivially negative, the sequence of sets $\{{\rm
SDS}^{(m)}(f)\}_{m=1}^{\infty}$ is called negatively terminating.}

\th{Definition 3.8.}\ {\it Given a form $f$, the sequence of sets
$\{{\rm SDS}^{(m)}(f)\}_{m=1}^{\infty}$ is neither positively
terminating nor negatively terminating, then it is called not
terminating.}

\th{Lemma 3.3.}\ {\it Given the form $f\in
\mathcal{R}[x_1,x_2,\cdots, x_n]$,\ and an arbitrary natural number
$m$, then the following equivalence relations hold.}

$1^\circ$ $f\in {\rm PSD} \Longleftrightarrow {\rm
SDS}^{(m)}(f)\subset {\rm PSD} $.

$2^\circ$ $f\notin \ {\rm PSD} \Longleftrightarrow \exists g\in {\rm
SDS}^{(m)}(f),g\notin \  {\rm PSD} $.

\th{proof}\ $1^\circ$
\begin{eqnarray*}
&&f\in {\rm PSD} \Longleftrightarrow f(X)\geq 0,\forall X\in
\Delta_n\\
&&\qquad \qquad \ \Longleftrightarrow f(X)\geq 0,\forall X\in
\bigcup \limits_{\sigma_m\in S_n}\cdots \bigcup \limits_{\sigma_2\in
S_n} \bigcup \limits_{\sigma_1\in S_n}(\Delta_n
B_{\sigma_1}B_{\sigma_2}\cdots B_{\sigma_m})\\
&&\qquad \qquad \ \Longleftrightarrow
f(B_{\sigma_1}B_{\sigma_2}\cdots B_{\sigma_m}X)\geq 0, \forall X\in
\Delta_n, \ \forall
B_{\sigma_i}\in {\rm PT_n} \\
&&\qquad \qquad \ \Longleftrightarrow g\in {\rm PSD}, \forall g\in
{\rm SDS}^{(m)}
(f)\\
&&\qquad \qquad \ \Longleftrightarrow   {\rm SDS}^{(m)}(f)\subset
{\rm PSD} .
\end{eqnarray*}

$2^\circ$ getting by $1^\circ$.

\sec{4\quad Proof of the Main Results}

\th{Theorem 1.}\ {\it Let $f\in \R[x_1,x_2,\cdots,x_n]$ be a form,
then there are

 $1^\circ$ $f\in ${\rm PSD} iff the
sequence of SDS sets of $f$ is positively terminating or not
terminating.

$2^\circ$ $f\notin $ PSD iff the sequence of SDS sets
 of $f$ is negatively terminating.

$3^\circ$  If $f$ is positive definite on $\R^n_+$, then its
sequence of SDS is positively terminating.}

Next we will show the proof of Theorem  step by step. First prove
$1^\circ$ and $2^\circ$.

\subsec{4.1\ Proof of $1^\circ$ and $2^\circ$ in Theorem 1}

\th{Proof.}\ By the definitions of the termination of sequence of
sets (definition 3.6,\ 3.7,\ 3.8) we know that $1^\circ$ is
equivalent to $2^\circ$. So we just need to prove $2^\circ$.

The proof of sufficiency is straightforward by the conclusion of
$2^\circ$ in Lemma 3.6. So to prove necessity, we will discuss the
problem on the standard simplex $\Delta_n$ for $f$ is homogeneous.

Suppose that the form $f\notin$ PSD. Then by the definition 1.1,\
$\exists X_0\in \Delta_n$ such that $f(X_0)<0$. And we suppose that
$X_0$ is interior to $\Delta_n$ for such point is obviously
existing.

Since polynomials are continuous on $\Delta_n$, there exists a
sufficient small spherical neighborhood $O(X_0,\epsilon)$ of $X_0$
satisfying $$\forall\  X \in O(X_0,\epsilon), \ f(X)<0.$$

By Lemma 3.3, we can choose the simplex that include the point $X_0$
one by one:
$$\Delta_n B_{\sigma_1},\ \Delta_n B_{\sigma_1}B_{\sigma_2},
\ \cdots, \ \Delta_n B_{\sigma_1}B_{\sigma_2}\cdots B_{\sigma_m},\
\cdots.$$

Then we obtain a sequence of nested closed sets
$$\Delta_n B_{\sigma_1} \supset \Delta_n B_{\sigma_1}B_{\sigma_2}
 \supset \cdots \supset \Delta_n B_{\sigma_1}B_{\sigma_2}\cdots B_{\sigma_m} \supset \cdots .$$

By $2^\circ$ of Lemma 3.2, the diameters of these bounded closed
sets are monotonically decreasing and tending to 0. According as the
theorem of nested closed set, $X_0$ is the only common point of the
above sequence of closed sets. That is, when $m\rightarrow +\infty$,
there exists a natural number $L$ such that
$$O(X_0,\epsilon)\supset \Delta_n B_{\sigma_1}B_{\sigma_2}\cdots B_{\sigma_L} \supset \cdots .$$
Thus
$$\forall\  X \in \Delta_n B_{\sigma_1}B_{\sigma_2}\cdots B_{\sigma_L}, \ f(X)<0.$$

Let $g=f(B_{\sigma_1}B_{\sigma_2}\cdots B_{\sigma_L}X),\
Y_0=(1,1,\cdots,1)^T$. Then we have
\begin{displaymath}
g(Y_0)=f(B_{\sigma_1}B_{\sigma_2}\cdots B_{\sigma_L}Y_0)<0.
\end{displaymath}
That is the form $f(B_{\sigma_1}B_{\sigma_2}\cdots B_{\sigma_L}X)$
is trivially negative. Then $\{{\rm SDS}^{(m)}(f)\}_{m=1}^{\infty}$
is negatively terminating.

The proof for $1^\circ$ and $2^\circ$ in Theorem 1 is completed. Now
we'll show the proof of $3^\circ$.

\subsec{4.2\ Proof of $3^\circ$ in Theorem 1}

First we'll give several Lemmas for the proof of $3^\circ$.

\th{Lemma 4.1}\ {\it Given a form of degree $d$
$$
g=\sum\limits_{|\alpha|=d}C_{\alpha}x^{\alpha_1}_1
x^{\alpha_2}_2\cdots x^{\alpha_n}_n.
$$
 If there is $g(X)=0$
at each point $X$ of the standard simplex $\Delta_n$, then $g$ is a
0 polynomial. That is, for all of the $\alpha$, we have
$C_{\alpha}=0$.}

Lemma 4.1 is a well known conclusion (see [8]), which can be proved
by induction.

\th{Lemma 4.2.}\ {\it Given a form of degree $d$
$$
f_m=\sum\limits_{|\alpha|=d}C_{(\alpha,m)}x^{\alpha_1}_1
x^{\alpha_2}_2\cdots x^{\alpha_n}_n.
$$
If the sequence $\{f_m\}_{m=1}^{\infty}$ uniformly converges to 0 on
the standard simplex $\Delta_n$, then the corresponding sequence of
coefficients $\{C_{(\alpha,m)}\}_{m=1}^{\infty}$ will converge to
0.}

\th{Proof.}\ We only give the proof for the binary form with degree
$d$, and the multivariate form can be gotten by induction.

Let
$$
f_m=C_{(d,0,m)}x_1^{d}+C_{(0,d,m)}x_2^{d}+x_1x_2(C_{(d-1,1,m)}x_1^{d-2}
+\cdots+C_{(1,d-1,m)}x_2^{d-2}).
$$

Then pick $(x_1,x_2)=(1,0),(0,1)$ and we have$\lim
\limits_{m\rightarrow \infty}C_{(d,0,m)}=0$, $\lim
\limits_{m\rightarrow \infty}C_{(0,d,m)}=0$.

Consider the new polynomial
$$
\begin{array}{ll}
f^{(1)}_m=f_m-(C_{(d,0,m)}x_1^{d}+C_{(0,d,m)}x_2^{d})\\
\qquad
=x_1x_2(C_{(d-1,1,m)}x_1^{d-2}+\cdots+C_{(1,d-1,m)}x_2^{d-2}).
\end{array}
$$
It is obviously that $\{f^{(1)}_m\}_{m=1}^{\infty}$ still uniformly
converges to 0 on the standard simplex $\Delta_2$, that is, for all
$\varepsilon>0$, when $m$ is sufficient large,
$$|x_1x_2(C_{(d-1,1,m)}x_1^{d-2}+\cdots+C_{(1,d-1,m)}x_2^{d-2})|<\varepsilon$$
holds.

Let $x_1=(1-\sqrt{\varepsilon}),x_2=\sqrt{\varepsilon}$,
substituting them in the above inequality, we have
$$|(1-\sqrt{\varepsilon})(C_{(d-1,1,m)}(1-\sqrt{\varepsilon})^{d-2}+\cdots+
C_{(1,d-1,m)}(\sqrt{\varepsilon})^{d-2})|<\sqrt{\varepsilon}.$$

And let $\varepsilon\rightarrow 0$, then we have $\lim
\limits_{m\rightarrow \infty}C_{(d-1,1,m)}=0$. Analogously, let
$x_1=\sqrt{\varepsilon},x_2=(1-\sqrt{\varepsilon})$, then we have
$\lim \limits_{m\rightarrow \infty}C_{(1,d-1,m)}=0$.

Do repeated discussion, and finally we have that every sequence
$\{C_{(\alpha,m)}\}_{m=1}^{\infty}$ converges to 0.

\th{Lemma 4.3.}\ {\it Given two forms of degree $d$
\begin{eqnarray*}
&&f_m=\sum\limits_{|\alpha|=d}C_{(\alpha,m)}x^{\alpha_1}_1
x^{\alpha_2}_2\cdots x^{\alpha_n}_n,\\
&&g=\sum\limits_{|\alpha|=d}C_{\alpha}x^{\alpha_1}_1
x^{\alpha_2}_2\cdots x^{\alpha_n}_n.
\end{eqnarray*}
If the sequence $\{f_m\}_{m=1}^{\infty}$ uniformly converges to $g$
on the standard simplex $\Delta_n$, then the corresponding sequence
of coefficients $\{C_{(\alpha,m)}\}_{m=1}^{\infty}$ will converge to
$C_{\alpha}$.}

\pf{Proof of $3^\circ$ in Theorem 1.}\ Consider the infinite
sequence of matrices, where $B_{\sigma_i}\in {\rm PT}_n$
$$B_{\sigma_1},B_{\sigma_2},\ \cdots,B_{\sigma_m},\cdots.$$
By Main Lemma, and let
$$B=\prod \limits_{m=1}^{\infty} B_{\sigma_m}=
\left[
\begin{array}{cccc}
b_1 & b_1 & \cdots & b_1 \\
b_2 & b_2 & \cdots & b_2 \\
\vdots& \vdots &\vdots &\vdots \\
b_n & b_n &\cdots & b_n
\end{array}
\right].$$ We denote the degree of form $f$ as $d$,
$X=(x_1,x_2,\cdots,x_n)^T$, then there always be
$f(b_1,b_2,\cdots,\allowbreak b_n)>0$ for $f$ is positive definite
on $\R^n_+$. Thus by the homogeneous of $f$, we know that
\begin{equation}
f(\prod \limits_{m=1}^{\infty} B_{\sigma_m}X)=f(BX)=
f(b_1,b_2,\cdots,b_n)(x_1+x_2+\cdots+x_n)^d
\end{equation}
is trivially positive.

Consider the infinite sequence of polynomials $\Sigma$
\begin{eqnarray}
\Sigma:\quad  f(X),\  f(B_{\sigma_1}X),\
f(B_{\sigma_1}B_{\sigma_2}X),\ \cdots, \
f(B_{\sigma_1}B_{\sigma_2}\cdots B_{\sigma_m}X) ,\ \cdots.
\end{eqnarray}

Next we will prove that there exists a positive integer $m_0$, for
$m>m_0$ such that $f(B_{\sigma_1}B_{\sigma_2}\allowbreak \cdots
B_{\sigma_m}X)$ is trivially positive.

On the one hand, for $f$ is continuous on $\Delta_n$, namely, for
all $\varepsilon>0$, there exists a spherical neighborhood
$O(P,\delta)\subset\Delta_n$ of $P=(b_1,b_2,\cdots,b_n)$ (If $P$ is
on the boundary of $\Delta_n$, then we take $O(P,\delta)\cap
\Delta_n$)) satisfying $\forall X_1,X_2\in O(P,\delta),
|f(X_1)-f(X_2)|<\varepsilon.$

On the other hand, by $2^\circ$ of Lemma 3.2, we know that the
diameters of following nested closed sets are monotonically
decreasing and tending to 0,
$$\Delta_n \supset \Delta_n B_{\sigma_1} \supset \Delta_n B_{\sigma_1}B_{\sigma_2}
 \supset \cdots \supset \Delta_n B_{\sigma_1}B_{\sigma_1}\cdots B_{\sigma_m} \supset \cdots$$
According as the theorem of nested closed set, these sets have only
a common point. Obviously it is $P=(b_1,b_2,\cdots,b_n)$.

Putting together the above two aspects, we have that when
$m\rightarrow +\infty$, there exists a natural number $L$ such that
$$
 O(P,\delta)\supset \Delta_n B_{\sigma_1}B_{\sigma_1}\cdots B_{\sigma_L} \supset \cdots.
$$
That is, when $\forall X_1\in \Delta_n
B_{\sigma_1}B_{\sigma_2}\cdots B_{\sigma_{m_1}}(={\rm
Im}(\psi_{\sigma_1}\psi_{\sigma_2}\cdots \psi_{\sigma_{m_1}}))$,\
$\forall X_2\in \Delta_n B_{\sigma_1}B_{\sigma_2}\cdots\allowbreak
B_{\sigma_{m_2}} (={\rm Im}(\psi_{\sigma_1}\psi_{\sigma_2}\cdots
\psi_{\sigma_{m_2}}))$  there is
$$|f(X_1)-f(X_2)|< \varepsilon,$$
namely,\ $\forall X\in \Delta_n$ there is
$$|f(B_{\sigma_1}B_{\sigma_2}\cdots B_{\sigma_{m_1}}X)-
f(B_{\sigma_1}B_{\sigma_2}\cdots B_{\sigma_{m_2}}X)|< \varepsilon$$

So the sequence of polynomials (7) uniformly converges to $f(BX)$ on
$\Delta_n$. By Lemma 4.3, the sequence of coefficients of the same
monomials in the sequence (7) will converge to the coefficients of
the same monomial of $f(BX)$. And the coefficients of $f(BX)$ are
all positive (see (6)), so there exists a positive integer $m_0$,
for $m>m_0$ such that $f(B_{\sigma_1}B_{\sigma_2}\cdots
B_{\sigma_m}X)$ is trivially positive. This minimum $m_0$ is called
index of trivially positive of sequence $\Sigma$.

Consider the ordered pairs $(\Sigma, k)$, where $k$ is the index of
trivially positive of sequence $\Sigma$. We still need to prove that
there is a $K\in \N$, such that $\forall (\Sigma, k), k\leq K$.

We are able to select
\begin{eqnarray}
(\Sigma_1, k_1),(\Sigma_2, k_2),\cdots,(\Sigma_t, k_t),\cdots
\end{eqnarray}
in turn, such that $k_1<k_2<\cdots<k_t<\cdots.$ If there is no
positive integer K satisfying $k\leq K$ for $\forall (\Sigma, k)$.
Then the sequence (8) can be infinite. That is, there is a $\Sigma$
which has no finite trivially  positive index. Contradiction.

So everyone of the set ${\rm SDS}^{(k)}(f)$ is trivially positive,
namely, ${\rm SDS}^{(k)}(f)$ is positively terminating.

\subsec{4.1\ Corollary, algorithm and other consideration}

It's well-known to us that if the binary form $f$ is a positive
semi-definite form with no square factors, then $f$ is definitely a
positive definite form.

\th{Corollary 4.4}\ {\it The binary form $f$ has no square factors,
then $f$ is positive semi-definite iff the sequence of sets of $f$
is positively terminating. And $f$ is not a semi-definite form iff
the sequence of sets of $f$ is negatively terminating.}

Thus we obtain the following algorithm by the theorem 1, which is
used to decide the positive semi-definite form.

\th{Algorithm(NEWTSDS)}

Input: form $f\in \Q[x_1,x_2,\cdots,x_n]$.\ \

Output: the form $f\in {\rm PSD}$,\ or $f\notin {\rm PSD}$.

T1: Let $F=\{f \}$.

T2: For each element in $F$ Compute $\bigcup \limits_{g\in F} {\rm
SDS}(g)$.

\quad\ \ \ Let $Temp=\{h|h\in\bigcup \limits_{g\in F} {\rm SDS}(g),
\hbox{$h$\ is\ not trivially\ positive}\}$.

\qquad \qquad  T21: If $Temp$ is empty set, then output $f\in$PSD
and terminate.

\qquad \qquad  T22: If there are trivially negative polynomials in
$Temp$, then output

\qquad \qquad  \qquad \ $f\notin {\rm PSD}$ and terminate.

\qquad \qquad  T23: Else, let $F=Temp$, go to step T2.

According to the above algorithm, we designed a Maple program called
TSDS3. It mainly has two orders, TSDS and NEWTSDS. TSDS chooses
$A_n$ as basic substitution matrix, and NEWTSDS chooses $T_n$ as
basic substitution matrix.

The order TSDS sometimes does not halt to positive definite
forms(PD). For example, $f=(3x_1+x_2-x_3)^2+\frac{x_3^2}{3}.$
However, the former two results of Theorem 1 are still hold for the
difference substitution based on $A_n$(Seeing [9]).

And to the order NEWTSDS, by Theorem 1, we confirm that the program
will terminate if the input form $f$ itself is not PSD, and judge
that $f\geq 0$ does not hold. Otherwise, if the input form $f$
itself is PSD, then the program either terminate(that is, judge that
$f\geq 0$ holds) or does not(namely, can't decide whether $f\geq 0$
holds). However, as the $3^\circ$ in Theorem 1 says, the algorithm
NEWTSDS is complete for the positive definite form. And if we enter
the subprogram of eliminating square factors, then the algorithm
NEWTSDS is also complete for the decision of binary form.

We can generalize the Theorem 1 as follows

Let $R_1,R_2,\cdots,R_k$ be $n\times n$ column stochastic mean
matrices, such that
$$[P_1,P_2,\cdots,P_n]=\bigcup
\limits^{k}_{i=1}[P_1,P_2,\cdots,P_n]R_i,$$ where
$[P_1,P_2,\cdots,P_n]$ is arbitrarily a simplex in $\Delta_n$. And
the set SDS$^{(m)}(f)$ is definited as
$${\rm SDS}^{(m)}(f)=\bigcup\limits^{k}_{i_1=1}
\cdots \bigcup\limits^{k}_{i_m=1}f(R_{i_1}\cdots R_{i_m} X).$$ Then
the results of Theorem 1 still hold.

In addition, from the aspect of the method of partial cylindrical
algebraic decomposition (PCAD)$^{[10][11]}$, it is easily to prove
that:

\th{\bf Lemma 4.5.}\ {\it Given the form $f\in \R[x_1,x_2,\cdots,
x_n].$ There exists a positive integer $m$ such that the point set
$$K_m=\bigcup \limits_{\sigma_m\in S_n}\cdots
\bigcup \limits_{\sigma_2\in S_n} \bigcup \limits_{\sigma_1\in
S_n}B_{\sigma_1}B_{\sigma_2}\cdots
B_{\sigma_m}(\frac{1}{n},\cdots,\frac{1}{n})^T.$$ has at least one
sample point in each cell} (here all the lower dimensional cells are
discarded).

Lemma 4.5 gives another way for the completion of the algorithm
NEWTSDS. That is, to compute the upper bound of $m$, which is a fine
subject for further study.

\sec{5\quad Comparison between Theorem 1 and P\`olya's Theorem and
Applying examples}

\th{Theorem (P\`olya's)}\ {\it If a form $f\in
\R[x_1,x_2,\cdots,x_n]$ is positive on $\R^n_+$ when
$x_1+x_2+\cdots+x_n>0$, then for sufficiently large $N$ all the
coefficients of $$(x_1+x_2+\cdots+x_n)^Nf$$ are positive.}

P\`olya's Theorem was published in 1928 (in German) and is also in
$\ll$Inequalities$\gg^{[12]}$. From then on, this elegant and
beautiful result has many applications. In 1946, Habicht$^{[13]}$
skillfully used P\`olya's Theorem to give a constructive proof of a
positive definite case of Hilbert's 17th Problem. And recently ,
Schweighofer used P\`olya's Theorem to give an algorithm proof of
Schm$\ddot{u}$dgen's Positivstellensatz$^{[14][15]}$. Furthermore,
the quantitative form of P\`olya's Theorem has been used in the
study of some other problems$^{[16]}$, and the study of quantitative
problems themselves obtains very good results $^{[17][18]}$.

Next we will show the comparison between Theorem 1 and P\`olya's
Theorem in several following aspects:

\begin{center}{{\bf Table 1 \\ }}
{{\small
\begin{tabular}{|c|c|c|}\hline
{The given classes}& \  P\`olya's Theorem \ &  Theorem 1  \\
\hline
{PD}&  Completely decide & Completely decide  \\
\hline
{PSD with boundary zero}&  Partially apply & Partially apply  \\
\hline
{PSD with interior zero}& Can't decide & Partially apply  \\

\hline {not PSD }& Can't decide & Completely decide \\
\hline
\end{tabular}}}
\end{center}\quad

Next we will give some practical examples.

\th{Example 2.}(PD, A-G) \ \ Let $(x_1,x_2,\cdots, x_n)\in \R^n_+,$
and $\varepsilon>0$ is small enough. Consider the positive definite
form
$$
f=\sum \limits_{i=1}^n x_i^n-(n-\varepsilon )\prod \limits_{i=1}^n
x_i.
$$
In the book [12], this example is analyzed by P\`olya's method,
where when $N$ equals $\frac{n^3(n-1)}{2\varepsilon}$, we have a
trivially positive polynomial.

Analogously, the Hurwitz identity relation is recorded in the same
book,
\begin{eqnarray*}
&f_1&=\sum \limits_{i=1}^n x_i^n-n\prod \limits_{i=1}^n x_i\\
&&=\frac{1}{2n!}(\sum !(x_1^{n-1}-x_2^{n-1})(x_1-x_2)
+\sum !(x_1^{n-2}-x_2^{n-2})(x_1-x_2)x_3\\
&&\quad +\sum !(x_1^{n-3}-x_2^{n-3})(x_1-x_2)x_3x_4+\cdots),
\end{eqnarray*}
where the notion $\sum !$ denotes the sum of all permutations of the
variables. From the Hurwitz identity relation we easily get that the
elements of the set ${\rm SDS}(f_1)$ is trivially positive(the basic
substitution matrix can be any one of $A_n$ and $T_n$).

\th{Example 3$^{[19]}$} (not PSD, Vasc conjecture ) Try to decide
whether the inequality holds.
\begin{eqnarray*}
\frac{a_1-a_2}{a_2+a_3}+\frac{a_2-a_3}{a_3+a_4}+\frac{a_3-a_4}{a_4+a_5}+\frac{a_4-a_5}{a_5+a_6}
+\frac{a_5-a_6}{a_6+a_1}+\frac{a_6-a_1}{a_1+a_2}\geq 0,
\end{eqnarray*}
in which $a_i>0(i=1,\cdots,6)$.

Take off denominators of the left polynomial, and then denote the
new polynomial by $f$. Utilizing the program TSDS and executing
order $tsds(f)$, we have a counterexample: $a_1=84,a_2=7,
a_3=79,a_4=5,a_5=76,a_6=1.$ Analogously, it is easy to prove that
the inequalities like above in 3,4,5,7 variables(in case of 7 needs
longer time) still holds.

There are more examples in [1][2][19][20],we do not list them one by
one.

\vspace{3mm}
\th{Acknowledgements}
The author thanks referees a lot for their valuable opinions and
suggestions.
\vskip0.1in \no {\normalsize \bf References}
\vskip0.1in\parskip=0mm \baselineskip 15pt
\renewcommand{\baselinestretch}{1.15}

\footnotesize
\parindent=6mm

\REF{1\ }Yang L. Solving Harder Problems with Lesser Mathematics.
Proceedings of the 10th Asian Technology
 Conference in Mathematics, ATCM Inc, 37--46, 2005

\REF{2\ } Yang L. Difference substitution and automated inequality
proving. {\it Journal of Guangzhou University}(Natural Science
Edition), {\bf 5(2)}: 1--7 (2006)

\REF{3\ } Yang L. Some new advances of automated inequality proving.
Summer School on Symbolic Computation. Beijing: China. July 15--22,
2006

\REF{4\ } Cohn H. Products of stochastic matrices and applications.
{\it Internat. J. Math. $\&$ Math. Sci.}, {\bf 12}: 209--233 (1989)

\REF{5\ } Leizarowitz A. On infinite products of stochastic
matrices. {\it Lin. Alg. and its Applications}, {\bf 168}: 189-219
(1992)

\REF{6\ } Armstrong M A. Basic Topology. New York (Berlin,
Heidelberg): Springer-Verlag, 125--127, 1983

\REF{7\ } Basu S, Pollack R, Roy M F. Algorithms in Real Algebraic
Geometry (2nd). New York (Berlin, Heidelberg): Springer-Verlag,
215--217, 2006

\REF{8\ } Cox D, Little J, O'Shea D. Ideals, Varieties, and
Algorithms. New York (Berlin, Heidelberg): Spring-Verlag, 3--4 1996

\REF{9\ } Yang L, Yao Y. Difference substitution matrices and
decision on nonnegativity of polynomials. {\it Journal of Systems
Science and Mathematical Sciences}, {\bf 29(9)}: (2009)

\REF{10\ } Collins G E, Hong H. Partial cylindrical algebraic
decomposition for quantifier elimination, {\it Journal of Symbolic
Computation}, {\bf 12(2)}: 299--328 (1991)

\REF{11\ } Yang L, Xia B C. Computational real algebraic geometry.
In: Wang D M {\it et al eds.} Selected lecture in symbolic
compution. Beijing: Tusinghua Univ. Press, 100-149, 2003

\REF{12\ } Hardy G H, Littlewood J E, P\`olya G. Inequalities (2nd).
Cambridge: Camb.Univ.Press, 1952

 \REF{13\ } Habicht W. $\ddot{U}$ber
die Zerlegung strikte definter formen in quadrate. {\it coment Math.
 Helv}, {\bf 12}: 317--322 (1940)

\REF{14\ } Schweighofer M. An algorithmic approach to Schm\`udgen's
Positivstellensotz. {\it J. Pure and Appl.Alg}, {\bf 166}: 307--319
( 2002)

\REF{15\ } Schweighofer M. On the complexity of Schm\`udgen's
Positivstellensotz. {\it J. Complexity}, {\bf 20}: 529--543 (2004)

\REF{16\ } de Klerk E, Pasechnik D. Approximation of the stability
number of a graph via copositive
 programming. {\it SIAM J. Optimization}, {\bf 12}: 875--892 (2002)

\REF{17\ } Powers V, Reznick B. A new bound for P\`olya's Theorem
with applications to polynomials positive on polyhedra. {\it J. Pure
Appl. Alg}, {\bf 164}: 221--229 (2001)

\REF{18\ } Powers V, Reznick B. A quantitative P\`olya's Theorem
with corner zeros. In: Proc. ISSAC'06. New York: ACM Press, 2006

\REF{19\ } Yang L, Xia B C. Automated Proving and Discoverering on
Inequalities. Beijing: Science Press, 174, 2008

\REF{20\ } Xu J, Yao Y. Rationalizing algorithm and automated
proving for a class of inequalities involving radicals. {it Chinese
Journal of computers}, {\bf 31(1)}: 24--31 (2008)




\end{document}